\documentclass[%
 reprint,
superscriptaddress,
nofootinbib,
 amsmath,amssymb,
 aps,
pra,
]{revtex4-2}

\usepackage{graphicx}
\usepackage{dcolumn}
\usepackage{bm}
\usepackage[dvipsnames]{xcolor}
\usepackage[colorlinks=true]{hyperref}
\hypersetup{
    colorlinks=true,
    linkcolor=Green,
    citecolor=blue,
    filecolor=magenta,      
    urlcolor=cyan,
    pdftitle={Shell-model study of 28Si: coexistence of oblate, prolate and superdeformed shapes},
    }

\begin{document}

\preprint{APS/123-QED}

\title{Shell-model study of $^{28}$Si: coexistence of oblate, prolate and superdeformed shapes}

\author{Dorian~Frycz}
\email{dorianfrycz@fqa.ub.edu}
\affiliation{%
 Departament de F\'isica Qu\`antica i Astrof\'isica, Universitat de Barcelona, 08028 Barcelona, Spain.
}
\affiliation{
Institut de Ci\`encies del Cosmos, Universitat de Barcelona, 08028 Barcelona, Spain.
}%
\author{Javier~Men\'endez}%
\email{menendez@fqa.ub.edu}
\affiliation{%
 Departament de F\'isica Qu\`antica i Astrof\'isica, Universitat de Barcelona, 08028 Barcelona, Spain.
}%
\affiliation{
Institut de Ci\`encies del Cosmos, Universitat de Barcelona, 08028 Barcelona, Spain.
}
\author{Arnau~Rios}%
\email{arnau.rios@fqa.ub.edu}
\affiliation{%
 Departament de F\'isica Qu\`antica i Astrof\'isica, Universitat de Barcelona, 08028 Barcelona, Spain.
}%
\affiliation{
Institut de Ci\`encies del Cosmos, Universitat de Barcelona, 08028 Barcelona, Spain.
}
\author{Benjamin~Bally}
\email{benjamin.bally@cea.fr}
\affiliation{%
ESNT, IRFU, CEA, Universit\'e Paris-Saclay, 91191 Gif-sur-Yvette, France
}
\author{Tom\'as~R.~Rodr\'iguez}
\email{tomasrro@ucm.es}
\affiliation{Grupo de F\'isica Nuclear, Departamento EMFTEL and IPARCOS, Universidad Complutense de Madrid, 28040 Madrid, Spain}
\author{Antonio~M.~Romero}%
\email{antonio.marquezromero@fujitsu.com}
\affiliation{%
 Departament de F\'isica Qu\`antica i Astrof\'isica, Universitat de Barcelona, 08028 Barcelona, Spain.
}%
\affiliation{
Institut de Ci\`encies del Cosmos, Universitat de Barcelona, 08028 Barcelona, Spain.
}

\date{\today}

\begin{abstract}
We study the shape coexistence in the nucleus $^{28}$Si with the nuclear shell model using numerical diagonalizations complemented with variational calculations based on the projected generator-coordinate method. The theoretical electric quadrupole moments and transitions as well as the collective wavefunctions indicate that the standard USDB interaction in the $sd$ shell describes well the ground-state oblate rotational band, but misses the experimental prolate band.
Guided by the quasi-SU(3) model, we show that the prolate band can be reproduced in the $sd$ shell by reducing the energy of the $0d_{3/2}$ orbital. \textcolor{black}{Alternatively, in the extended $sdpf$ configuration space a modification of the SDPF-NR interaction
that accommodates cross-shell excitations
also reproduces the oblate and prolate bands.} Finally, we address the possibility of superdeformation in $^{28}$Si within the $sdpf$ space. \textcolor{black}{Our results indicate that superdeformed structures appear at about $18$-$20$~MeV}.  
\end{abstract}

\maketitle

\section{\label{Introduction}Introduction}

The intricate character of nucleon-nucleon forces combined with the complex nature of quantum many-body systems leads to the emergence of a diverse array of collective structures in nuclei. Driven by the quadrupole-quadrupole component of the nuclear force, deformations are notably prevalent in the nuclear chart for nuclei away from magic numbers. Moreover, within a limited energy range of a few MeV, distinct collective structures can appear in the same nucleus, a phenomenon usually referred to as shape coexistence~\cite{heyde2011shape,Nowacki:2021fjw,Shape_coexistence_experimental_view}. For instance, medium-mass and heavy nuclei such as $^{16}$O \cite{Carter:1964zz,Chevallier:1967zz}, 
$^{40}$Ca \cite{Ideguchi:2001zz}, $^{56}$Ni \cite{Rudolph:1999zz} or $^{186}$Pb \cite{andreyev2000triplet}, among many-others~\cite{poves2016shape,otsuka2016role,Gade:2016xiy},
show well-established spherical and differently-deformed states at low energies.

\begin{figure*}[t]
\centering
\includegraphics[width=1.0\linewidth]{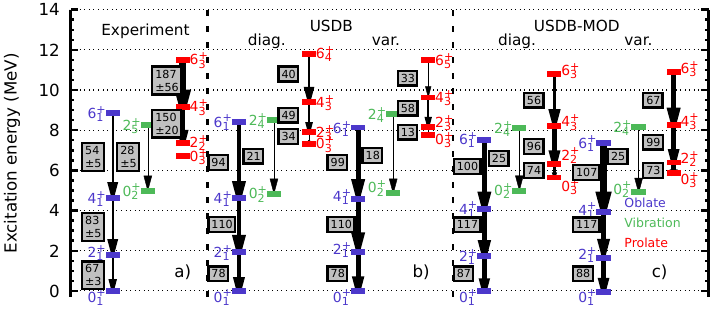}
\caption{{Band structure of the lowest-lying positive parity states of $^{28}$Si: a) experiment \cite{Si28exp}; 
b) results for the USDB interaction with diagonalization (left side) and the variational PGCM (right side); 
c) results for the USDB-MOD interaction with diagonalization (left) and the  PGCM (right). 
The arrows indicate inband $B(E2)$ transition strengths (in $e$$^2$ fm$^4$), with larger values associated to more deformed shapes.}}
\label{fig:Si28Spectrum_USDB}
\end{figure*}

The $^{28}$Si nucleus, with $Z=14$ protons and $N=14$ neutrons, fills exactly half of the $sd$ shell in the naive shell-model scheme \cite{Mayer,haxel1949magic}. Specifically, the $0d_{5/2}$ orbital ---using the conventional $nl_j$ notation where $n,l,j$ are the radial, orbital, and total angular momentum quantum numbers, respectively---is filled. This configuration leads to a spherical $0^+_\text{gs}$ ground state. However, experimental data  indicates the presence of a rotational band on top of the ground state with oblate deformation~\cite{Si28exp}. Alternatively, following Elliott's SU(3) framework~\cite{Elliot} based on the quadrupole-quadrupole interaction within the $sd$ shell, the $^{28}$Si ground state would exhibit a degenerate prolate/oblate deformation \cite{bernier196728si}. This is also in contrast to experiment, because the prolate rotational band emerges at an excitation energy of $\sim6.5$~MeV. Furthermore, the oblate ground state exhibits a $\beta$-vibration with $0^+_2$ bandhead at $\sim5$~MeV. Figure \ref{fig:Si28Spectrum_USDB} a) presents this coexistence of oblate and prolate collective structures. The difficulty to describe these with simple models underscores the complex nature of $^{28}$Si. 

Previous theoretical works have attempted at describing the shape coexistence in $^{28}$Si. While algebraic SU(3)-based approaches predict the main features of the oblate and prolate deformed bands~\cite{bernier196728si,vargas2001quasi}, early shell-model studies focus on the lowest-energy levels of the oblate ground-state band and its $\beta$-vibration~\cite{soyeur1972structure,wildenthal1973shell}. Likewise, to the best of our knowledge, more recent shell-model investigations using the phenomenological USDB~\cite{USDB} and several {\it ab initio} interactions based on the no-core shell model~\cite{Smirnova:2019yiq,abinitio-Si-P} and valence-space in-medium renormalization group (VS-IMSRG) approach~\cite{Li:2023bel} also limit their scope to the excitation spectrum and few electromagnetic transitions between oblate low-lying states.
In addition, $^{28}$Si has been recently studied with the antisymmetrized molecular dynamics approach~\cite{AMD} and the Hartree-Fock-Bogoliubov (HFB) plus quasiparticle random-phase approximation method~\cite{sd-triaxial}. In all these cases, a quality description of the measured nuclear structure of $^{28}$Si has been shown to be challenging.
First, the ground-state oblate rotational band does not behave as an ideal rotor, a feature typically not captured in these studies. Additionally, the deformation of the oblate band is either overestimated~\cite{Smirnova:2019yiq,abinitio-Si-P} or underestimated~\cite{AMD,sd-triaxial}. Finally, some works do not find a clear prolate band~\cite{abinitio-Si-P,Smirnova:2019yiq,sd-triaxial}, while in others its collectivity is much larger than in experiment~\cite{AMD}. 

Furthermore, Ref.~\cite{AMD} predicted the existence of a superdeformed shape in $^{28}$Si, spurring experimental efforts that so far have not found such extreme deformation~\cite{kubono1986highlySD,jenkins2012candidate,Si28_SD}. Superdeformed structures have been identified in several medium-mass $sdpf$-shell nuclei, such as $^{24}$Mg~ \cite{dowieMg24}, $^{36}$Ar~\cite{Ar36}, $^{40}$Ar~\cite{ideguchi2010superdeformation}, $^{40}$Ca~\cite{Ideguchi:2001zz} and $^{42}$Ca~\cite{Ca42}, which are generally well described by theoretical shell-model~\cite{Caurier:2005sq,Ca40,Ca42,Ar40_2} and antisymmetrized molecular dynamics~\cite{Chiba:2015zxa,Ar40} studies. As in $^{28}$Si, a so far unmeasured superdeformed band has been predicted in~$^{32}$S, in this case based on a projected HFB calculation~\cite{rodriguez2000S32}.

Overall, a unified description of all collective structures in $^{28}$Si presents a challenge for nuclear theory. We aim to gain insight into the shape coexistence of this nucleus guided by analytical models based on the SU(3) symmetry~\cite{Elliot,SU3} and employing state-of-the-art shell-model calculations. The latter include standard diagonalizations~\cite{Shellmodel} complemented with variational calculations based on beyond mean-field techniques~\cite{taurus,Taurus2}. In both approaches, we use standard phenomenological nuclear interactions tailored for the shell-model configuration space. 

The article is organized as follows. Section~\ref{sec:framework} introduces basic notions to characterize deformation in nuclei, and discusses the analytical SU(3)-based models and the numerical shell-model calculations performed in this work. The latter include both standard diagonalizations, outlined in Sec.~\ref{sec:diagonalization}, and variational calculations based on the projected-generator-coordinate method (PGCM), discussed in Sec.~\ref{sec:variational}. We present our theoretical results in Sec.~\ref{sec:results}, covering the oblate ground-state and $\beta$-vibration bands in Sec.~\ref{sec:oblate}, the prolate band in Sec.~\ref{sec:prolate}, and possible superdeformed states in Sec.~\ref{sec:superdeformed}. Finally, Sec.~\ref{sec:summary} summarizes our main results and provides an outlook for future work.

\section{Theoretical framework}
\label{sec:framework}

\subsection{Deformation in nuclei}

A prime signature of nuclear deformation is the appearance of structured bands due to the rotation of a permanently deformed shape in the intrinsic frame of reference. In the ideal rotor limit and assuming axial symmetry, these bands consist of levels with constant moment of inertia, $\mathcal{I}$, and energies related to its total angular momentum, $J$, such that $E_J=J(J+1)/(2\mathcal{I})$. For rotational bands with $J^{\pi}=0^+$ bandheads, where $\pi$ is the parity of the state, the levels can only adopt even values $J^{\pi}=0^+,2^+,4^+...$, due to symmetry restrictions \cite{Greiner}. 

The electric quadrupole moment also characterizes nuclear deformation. In the laboratory frame, it is defined by \cite{Suhonen}
\begin{eqnarray}
    Q_{\text{spec}}(J)=\sqrt{\frac{16\pi}{5}}\frac{1}{\sqrt{2J+1}}\langle JJ20 \vert JJ \rangle( J \vert \vert Q_{20} \vert \vert J)\,,
    \label{Qspec_def}
\end{eqnarray}
where the operator $Q_{20}=\sum_{i=1}^A e_i \,r^2_i \,Y_{20}(\theta_i,\phi_i)$ sums over the $A=N+Z$ nucleons in the nucleus and depends on the  charges $e_i$, the spherical harmonic $Y_{20}$, and the spherical coordinates of the nucleons: $r_i,\theta_i, \phi_i$. 
The double-bar $( J \vert \vert Q_{20} \vert \vert J)$ indicates a reduced matrix element and $\langle JJ20 \vert JJ \rangle$ is a Clebsch-Gordan coefficient in $\langle j_1 j_2 m_1 m_2 \vert j m \rangle$ notation, where $m$ is the projection of the total angular momentum 
\cite{edmonds1996angular}. The spectroscopic quadrupole moment is related to the intrinsic one by
\begin{eqnarray}
    Q_{0,s}=\frac{-(2J+3)}{J}Q_{\text{spec}}(J)\,. \label{Qspec}
\end{eqnarray}
A positive value of \textcolor{black}{the intrinsic quadrupole moment} represents a prolate shape while a negative one corresponds to an oblate shape.

\begin{figure}[b]
    \includegraphics[width=0.8\linewidth]{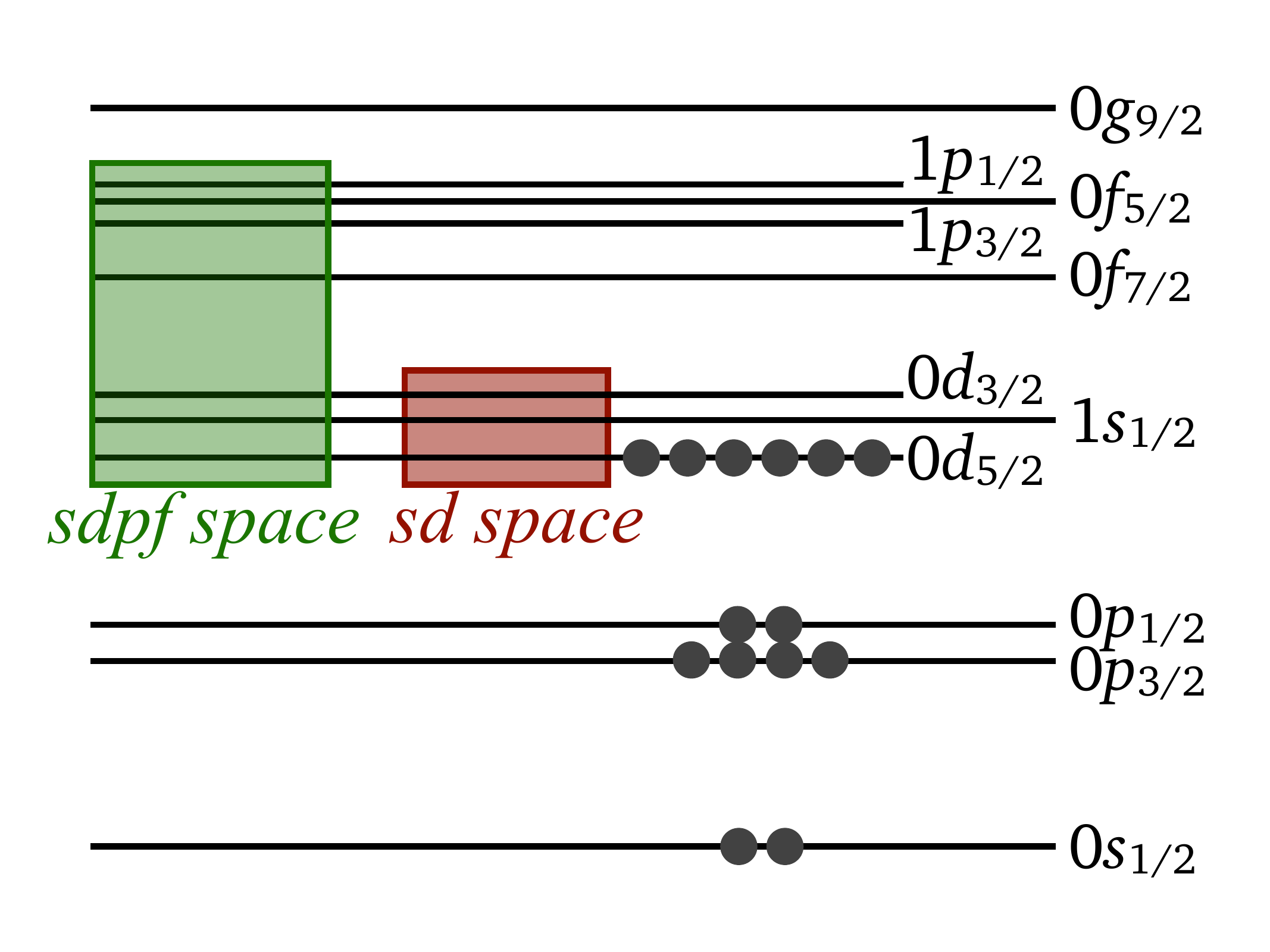}
    \caption{Shell model single-particle orbitals. Solid circles represent the naive shell-model filling for the $^{28}$Si ground state, and colored boxes highlight the two valence spaces considered in this work: the $sd$ shell ({red}) and the $sdpf$ space ({green}).}
    \label{fig:Spherical_levels}
\end{figure}

\begin{figure*}[t]
\centering
\includegraphics[width=0.8\linewidth]{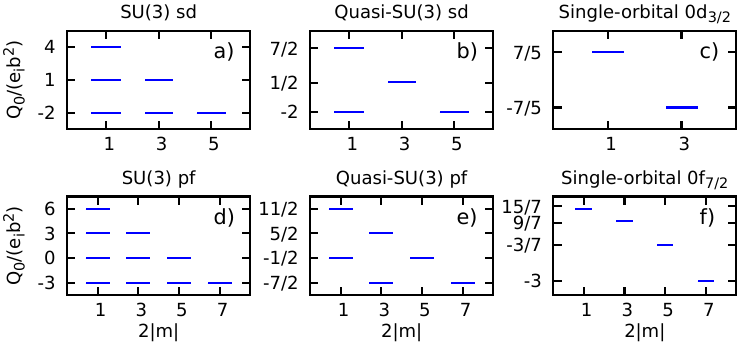}
\caption{{Quadrupole diagrams for the SU(3) variants considered in this work: SU(3) in the $sd$ (panel a) and $pf$ (d) shells, quasi-SU(3) for the $0d_{5/2}$-$1s_{1/2}$ (b) and $0f_{7/2}$-$1p_{3/2}$ + $0f_{5/2}$-$1p_{1/2}$ (e) orbital pairs, and the individual $0d_{3/2}$ (c) and $0f_{7/2}$ (f) single-particle orbitals. The dimensionless quadrupole moment $Q_{0}/(e_ib^2)$ is given for each 2$\vert m\vert$ value of the total angular momentum projection. Oblate states are obtained by filling the diagrams from below and prolate ones from
above.}}
\label{fig:SU(3)orbitals}
\end{figure*}

A complementary measure of the nuclear deformation comes from $B(E2)$ transition strengths:
\begin{eqnarray}
    B(E2;J_i\xrightarrow{} J_f)=\frac{1}{2J_i+1}\left( J_f\vert \vert Q_{20} \vert \vert J_i \right)^2\,,
\end{eqnarray}
where $J_i$ is the angular momentum of the initial state and  $J_f$ that of the final one.
For inband states, the $B(E2)$s are highly enhanced, while outband transition strengths are generally suppressed if the corresponding intrinsic states have different deformation. The intrinsic quadrupole moment, in the limit of large axial deformation, can also be extracted from inband $B(E2)$ transition strengths through the relation
\begin{eqnarray}
    Q_{0,t}=\pm \sqrt{\frac{16\pi B(E2,J\rightarrow J-2)}{5\vert \langle J200 \vert J-2\;0 \rangle \vert^{2} }}\,. \label{BE2} 
\end{eqnarray}
For a well-established rotational band, a common intrinsic quadrupole moment must emerge from static and transition quadrupole moments, $Q_{0,s}\approx Q_{0,t}$. 
\textcolor{black}{Moreover, $B(E2)$ transition strengths may be fragmented across several final states. For an initial state, we also consider the quadrupole obtained from the sum rule related to the expectation value of the squared quadrupole operator:
\begin{equation}
   \langle Q^2 \rangle =\frac{\sum_f\left( J_f\vert \vert Q_{20} \vert \vert J_i \right)^2}{2J_i+1},\quad Q_{0,\text{SR}}=\sqrt{{\frac{16\pi}{5}}{\langle Q^2\rangle }}. 
\end{equation}
}Thus, a rotational band is characterized by a sequence of excited levels with energies proportional to $J(J+1)$, connected by strong $B(E2)$ transitions with a consistent value of $Q_{\text{spec}}$ \textcolor{black}{which is close to the quadrupole sum rule value}. Nonetheless, the notion of nuclear shape has to be taken with caution as fluctuations of expectation values of quadrupole operators may prove significant \cite{Poves:2019byh}. 

\subsection{SU(3) and quasi-SU(3) models}
\label{sec:SU3}

In the naive shell model, the nucleus is bound through a spherical mean-field Hamiltonian \cite{Mayer}
\begin{equation}
    \mathcal{H}_0=\frac{{\vec{p}\,}^2}{2m}+\frac{{\vec{r}\,}^2}{2mb^4}+\mathcal{A}\,{\vec{l}\ }^2+\mathcal{B}\,\vec{l}\cdot\vec{s}\,, \label{mean_field_potential}
\end{equation}
which features a harmonic oscillator with length parameter $b$, complemented by orbital angular momentum ($\vec{l}$) and spin-orbit ($\vec{l}\cdot\vec{s}$) terms, weighted by coefficients $\mathcal{A}$ and $\mathcal{B}$.
Figure \ref{fig:Spherical_levels} illustrates that in this scheme, the 14 protons and 14 neutrons in $^{28}$Si occupy single-particle states up to filling the $0d_{5/2}$ orbital. This configuration represents a spherical state, not observed experimentally.

In the shell-model framework,  the deformation of a given nucleus can be accommodated within Elliott's SU(3) model \cite{Elliot}, which  considers a Hamiltonian without spin-orbit term ($\mathcal{B}=0$) restricted to a major shell with an attractive ($\chi>0$) quadrupole-quadrupole interaction:
\begin{eqnarray}
    \mathcal{H}=\mathcal{H}_0-\chi (Q_2\cdot Q_2)\,. \label{Elliot_H}
\end{eqnarray}
Thus, by maximizing the quadrupole moment, nuclei can lower their energy. 
For $^{28}$Si, in the naive shell model the $s$ and $p$ shells are full, leading to a spherical configuration with \textcolor{black}{a vanishing quadrupole moment.} Thus, 
the associated deformation for $^{28}$Si can be obtained by filling the quadrupole diagram for the $sd$ shell in Fig.~\ref{fig:SU(3)orbitals}~a), where each level---fourfold degenerated in spin and isospin projections---represents the contribution of a nucleon to the (dimensionless) quadrupole moment of the nucleus. Prolate shapes arise from filling levels from above and oblate ones from below.  The quadrupole moment of a given configuration is the sum of the individual nucleon contributions: \textcolor{black}{
\begin{eqnarray}
    Q_0=\sum_i (e_i\,Q_{0,i} \pm3\ \tilde e)\ b^2\,, \label{SU3_quadrupole_sum}
\end{eqnarray}
where we add  (subtract) $3\ \tilde e\ b^2$ units for prolate (oblate) shapes to match with ideal rotors~\cite{Nilson,Retamosa}. We choose the electric charges $e_n=0.46e$ for neutrons and $e_p=1.31e$ for protons \cite{Charges_Zuker_Dufour}, where $e$ is the elementary electric charge,  and we define the average charge of a nucleon as $\tilde{e}=(e_p+e_n)/2=0.88e$. }
The oscillator length is approximated by \cite{b2}
\begin{eqnarray}
    b^2 \simeq \frac{41.4}{45 A^{-1/3}-25 A^{-2/3}}\ \text{fm}^2\,. \label{b2_param}
\end{eqnarray}
For $^{28}$Si, the two fillings give the same value, \textcolor{black}{$\vert Q_0\vert=27\ \tilde e\ b^2$},
therefore predicting both shapes to be degenerate in energy.
In contrast, the experimental ground state is oblate and the prolate bandhead appears $\sim6.5$ MeV higher.

However, as mentioned above, the SU(3) model neglects the strong spin-orbit nuclear force. More complex SU(3)-based models have been suggested to account better for rotational bands in $sd$-shell nuclei~\cite{Zbikowski:2020srq}.
Here we use the quasi-SU(3) model~\cite{SU3}, which incorporates the spin-orbit splitting, and exploits the fact that the $\Delta j=2$ single-particle matrix elements of the $Q_2$ operator are much larger than those with $\Delta j=1$. The quasi-SU(3) model highlights the collectivity driven by $\Delta j=2$ orbitals, the $0d_{5/2}$-$1s_{1/2}$ doublet in the case of the $sd$ shell. Figure~\ref{fig:SU(3)orbitals}~b) shows the corresponding quadrupole diagram for this pair of orbitals.
The quasi-SU(3) scheme treats other single-particle orbitals separately, with $Q_{0}$ given by~\cite{Nilson}
\begin{equation}
    Q_0=\sum_m (2n+l+3/2)\frac{j(j+1)-3m^2}{2j(j+1)}\,e_i\,b^2\,.
\label{eq:Singleorbit}
\end{equation}
Figure~\ref{fig:SU(3)orbitals}~c) allows one to obtain the contribution of the $0d_{3/2}$ orbital to the quadrupole moment.

\begin{table}[b]
\centering
\caption{{$^{28}$Si quadrupole moments for the experimental oblate and prolate bands, from $B(E2)$ transition strengths (top row), compared to the predictions for $n$p-$n$h configurations in the quasi-SU(3) and SU(3) models in the $sd$ shell.}
}
\label{tab:npnhconfig_analytical}
\begin{tabular}{c c c} \hline \hline \\[-1em]
 & \multicolumn{2}{c}{\textcolor{black}{$Q_{0,t}$} ($e$ fm$^2$)} \\ \hline \\[-1em]
 & Oblate  &  Prolate \\ \hline \\[-1em]
Experiment~\cite{Si28exp} &    $\text{-57.3}\pm\text{0.7}$      &     $\text{72}\pm\text{7}$      \\ \hline \\[-1em]
Quasi-SU(3): $0d_{5/2}$-$1s_{1/2}$ + $0d_{3/2}$ &        \multicolumn{2}{c}{\textcolor{black}{$Q_{0}$  ($e$ fm$^2$)}}    \\ \hline 
$0$p-$0$h       &   $-51.4$      &    $33.3$    \\
$2$p-$2$h       &     $-62.9$     &     $53.9$   \\
$4$p-$4$h       &    $-74.4$      & $74.4$      \\
$6$p-$6$h       &   $-53.9$       &   $62.9$    \\
$8$p-$8$h       &  $-33.3$       &  $51.4$   \\  \hline  
SU(3): $sd$      &  $-81.7$       &   $81.7$       \\ \hline \hline
\end{tabular}
\end{table}

Table~\ref{tab:npnhconfig_analytical} lists the quadrupole moments, compared to experimental data, for different $0d_{5/2}$-$1s_{1/2}$ + $0d_{3/2}$ quasi-SU(3) configurations. 
The notation $n$p-$n$h denotes the promotion of $n$ nucleons from the $0d_{5/2}$-$1s_{1/2}$ orbitals to the $0d_{3/2}$ one. 
For example, we fill diagram \ref{fig:SU(3)orbitals} b)  with 12 nucleons from below to study the oblate $0$p-$0$h configuration, which leads to a quadrupole moment \textcolor{black}{$Q_{0}=-17\  \tilde e\ b^2=-51.4$~$e$~fm$^{2}$}.
In contrast, for a prolate $4$p-$4$h configuration, we fill diagrams~\ref{fig:SU(3)orbitals} b) and \ref{fig:SU(3)orbitals} c) with 8 and 4 nucleons from above, respectively, to reach
\textcolor{black}{$Q_{0}=
24.6\ \tilde e\ b^2=74.4$~$e$~fm$^{2}$}. The large $Q_{0}$ value for the oblate $0$p-$0$h configuration is remarkable, because just by 
populating the $1s_{1/2}$ orbital the nucleus gains much correlation energy with respect to the closed $0d_{5/2}$ spherical picture. In fact, this gain overcomes the energy difference between the $1s_{1/2}$ and $0d_{5/2}$ orbitals, so that the system gravitates towards an oblate deformed shape instead of the spherical one. \textcolor{black}{We note that Ref.~\cite{Utsuno:2012qf} provides an alternative argument to motivate the oblate shape of silicon based on the mixing between $0d_{5/2}$ and $1s_{1/2}$ orbital configurations.} The experimental value is \textcolor{black}{similar although somewhat larger than} the oblate quasi-SU(3) prediction: \textcolor{black}{$ Q_{0,t}=-57.3$~$e$~fm$^{2}$} 
\cite{Si28exp}, 
\textcolor{black}{suggesting that other $n$p-$n$h contributions are needed to achieve the experimental deformation}. 

The results of table~\ref{tab:npnhconfig_analytical} provide insights on the interplay between excitations and deformation. On the one hand, in the quasi-SU(3) scheme, the $0$p-$0$h prolate configuration is disfavored due to its reduced quadrupole moment. On the other hand, the experimental prolate band is predicted to be dominated by $4$p-$4$h configurations, because they show the largest $Q_0$ value. However, $2$p-$2$h states---which require less single-particle energy---may contribute as well, since the experimental $Q_{0,t}$ value is in between those of $2$p-$2$h and $4$p-$4$h configurations.

\subsection{Nuclear shell model} \label{Many-body methods}

Guided by the findings of Sec.~\ref{sec:SU3}, we use the nuclear shell model~\cite{Caurier:2005sq,Otsuka19,Brown01} to study  quantitatively the shape coexistence of differently deformed states in $^{28}$Si. Our shell-model calculations cover two alternative configuration spaces, shown in Fig.~\ref{fig:Spherical_levels}: the $sd$ shell, including the neutron and proton $0d_{5/2}$, $1s_{1/2}$, and $0d_{3/2}$ single-particle orbitals, and the $sdpf$ space, which additionally includes the $0f_{7/2}$, $1p_{3/2}$, $0f_{5/2}$, and  $1p_{1/2}$ orbitals. In both cases there is a $^{16}$O core. Thus, we are left with $N_v=Z_v=6$ valence neutrons and protons in the configuration space.

The nuclear many-body problem to solve reads
\begin{align}
         \mathcal{{H}}_{\text{eff}}\vert \Psi_{\text{eff}} \rangle=E\vert \Psi_{\text{eff}} \rangle\,, \label{H_eff}
\end{align}
where $\mathcal{{H}}_{\text{eff}}$ is the effective Hamiltonian suited for the configuration space. Here we use  USDB~\cite{USDB}, the interaction of choice in the $sd$ shell. We also introduce a slightly modified Hamiltonian in this space, USDB-MOD, as discussed later.
Moreover, we employ \textcolor{black}{a modification of the SDPF-NR~\cite{SDPF} interaction}, which gives a good description of neutron-rich Si isotopes in the $sdpf$ configuration space. In order to get additional insights and access larger configuration spaces than in usual shell-model studies, we complement standard shell-model diagonalizations with a variational approach based on the PGCM~\cite{taurus,Taurus2}.

\subsubsection{Diagonalization} \label{sec:diagonalization}

The standard solution of  the shell-model many-body problem involves the diagonalization of $\mathcal{{H}}_{\text{eff}}$ in the many-body basis of Slater determinants in the configuration space. Therefore, nuclear states are linear combinations,
\begin{align}
\vert\Psi_{\text{eff}} \rangle =&\sum_{i}a_{i}\left|\Phi_{i}\right\rangle\,,  
\end{align}
with amplitudes $a_i$, of Slater determinants, $\left|\Phi_{i}\right\rangle$,
\begin{align}
\left|\Phi_{i}\right\rangle =&\,c_{i1}^{\dagger}c_{i2}^{\dagger}\dots c_{iA_v}^{\dagger}\left|0\right\rangle  \,,
\label{eq:shellmodel_state}
\end{align}
where $\left|0\right\rangle$ is the bare vacuum,  and Slater determinants are built with one creation operator $c^{\dagger}_{l}$
---with corresponding annihilation operator $c_{l}$---
for each of the $A_v$ nucleons in the configuration space. These states have good quantum numbers $J^{\pi}$, according to the symmetries of $\mathcal{{H}}_{\text{eff}}$.

We perform diagonalizations in the $sd$ and $sdpf$ spaces using the Lanczos method through the ANTOINE code~\cite{Caurier:2005sq,FNowacki}. 
These results can be considered as an exact solution, since we impose that the eigenvalues of the Hamiltonian are converged to better than $0.5$~keV.
However, the $sdpf$ space leads to a configuration size of $8.2\cdot10^{11}$ Slater determinants for $^{28}$Si, which is beyond our diagonalization capabilities. In this space, we truncate the configurations considered in our diagonalizations, see Secs.~\ref{sec:prolate} and~\ref{sec:superdeformed} for details.

\subsubsection{Variational PGCM}
\label{sec:variational}

We complement our results with a variational solution of the nuclear shell model many-body problem based on beyond-mean-field methods. This framework is well suited for very large configuration spaces, as it finds an approximation to the exact nuclear state, and then restores the quantum symmetries broken at the mean-field level. Here we study the shape coexistence of $^{28}$Si with the PGCM using the Taurus suite \cite{taurus,Taurus2}, which has been previously applied to other medium-mass nuclei in good agreement with shell-model diagonalizations ~\cite{Bally:2019miu,sanchez2021variational,tesis}.
Similar approaches have been applied to medium-mass and heavy nuclei~\cite{shimizu2021generator,Kaneko:2021chs,Kisyov:2022rsx}, including the discrete non orthogonal shell model~\cite{DaoNowacki,ISOLDE:2023iwf,Nies:2023tfn}.

The PGCM uses a set of reference states $\{\phi\}$ to construct the many-body basis.
We choose the Bogoliubov quasiparticle states that are vacua of the Bogoliubov quasiparticle operators $\{\beta_k,\beta^{\dagger}_k\}$ defined through the unitary transformation:
\begin{align}
    {\beta}_k^{\dagger}&=\sum_l U_{lk}{c}^{\dagger}_l + V_{lk}{c}_l, \\
    {\beta}_k&=\sum_l U_{lk}^*{c}_l + V_{lk}^*{c}^{\dagger}_l,
\end{align}
where $U_{lk}$ and $V_{lk}$ are the variational parameters.
To ensure that particle number is conserved on average in the HFB states, 
we employ two Lagrange multipliers, $\lambda_N$ and $\lambda_Z$, in the Hamiltonian: 
\begin{eqnarray}
    \mathcal{{H}}_{\text{eff}}'=\mathcal{{H}}_{\text{eff}}-\lambda_Z O_{Z}-\lambda_N O_{N}-\sum_i\lambda_i{O}_i\,,
\end{eqnarray}
where $O_Z$ and $O_N$ are particle number operators for proton and neutron spaces, respectively. 
Other constraints are implemented through the additional operators, $O_{i}$, and Lagrange multipliers, $\lambda_i$.
More specifically, we also constrain the quadrupole moment operators, as we are interested in nuclear deformation. However, rather than constraining directly the \textcolor{black}{$Q_{2\mu}=r^2Y_{2\mu}(\theta,\phi)$} operators, we use the deformation parameters $\beta$ and $\gamma$, defined as \cite{taurus} \textcolor{black}{
\begin{align}
    \beta&=\frac{4\pi }{3R_0^{2}A}\frac{e_{\text{mass}}}{e}\sqrt{\langle \overline{Q}_{20}\rangle^2+2\langle\overline{Q}_{22}\rangle^2}, \\ 
    \gamma&=\text{arctan}\left( \frac{\sqrt{2}\langle\overline{Q}_{22}\rangle}{\langle \overline{Q}_{20}\rangle}\right), 
\end{align}
}where $\overline{Q}_{2\mu}=\left( {{Q}_{2\mu}+{Q}_{2-\mu}} \right)/{2}$ is the Hermitian average of the quadrupole moment operator, \textcolor{black}{$e_{\text{mass}}=e_p+e_n=1.77e$} and $R_0=1.2A^{1/3}$ fm represents the nuclear radius without deformation. Here, $\beta$ denotes the magnitude of the deformation and $\gamma$, its type~\cite{Suhonen}.  If $\overline{Q}_{21}=0$, which we impose as an additional constraint, $\gamma=0^{\circ}$ represents a prolate deformation, whereas $\gamma=60^{\circ}$ indicates an oblate shape. \textcolor{black}{When comparing to experimental data, we also use the second-order definition $\beta=\beta_2(1+0.36\beta_2)$ \cite{Shape_coexistence_experimental_view}.}

The variational parameters  $U_{lk}$ and $V_{lk}$ can be extracted by minimizing either the HFB energy or the variation after particle-number projection (VAP) energy:
\begin{eqnarray}
    E_{\text{VAP}}(\phi)=\frac{\langle \phi \vert H_{\text{eff}} P^{N} P^{Z} \vert \phi \rangle}{\langle \phi \vert P^{N} P^{Z} \vert \phi \rangle} -
    \sum_i \langle \phi \vert \lambda_i O_i \vert \phi \rangle\,,
\end{eqnarray}
where $P^{N}$ and $P^{Z}$ are the neutron and proton number projectors \cite{projection}. The VAP minimization, albeit computationally more demanding, provides wavefunctions with well defined proton and neutron numbers. This choice is variationally more general and captures more pairing correlations~\cite{rodriguez2005restricted}, which leads to lower energies, closer to the exact solution. In this work, we show results calculated with VAP reference states, although the HFB approach yields similar deformation properties. 
We consider Bogoliubov quasiparticle states that have (positive) parity symmetry and we do not consider proton-neutron mixing.

Next, we still need to project the reference states obtained from the minimization process on neutron and proton numbers~\cite{Bally:2019miu} 
and total angular momenta 
\begin{eqnarray}
    \vert \phi^{NZJ} \rangle={P}^{N} {P}^{Z}{P}^{J}_{MK} \vert \phi \rangle\,,
\end{eqnarray}
where ${P}^{J}_{MK}$ projects onto total angular momentum $J$ with third components $M$ and $K$ in the laboratory and intrinsic frames of reference, respectively \cite{projection}. 
Finally, we consider configuration mixing through the GCM: 
\begin{eqnarray}
   \vert \Psi^{NZJM}_{\sigma,\text{GCM}} \rangle = \sum_{qK} f_{\sigma;qK}^{NZJM} \vert \phi^{NZJ}(q)  \rangle\,,
\end{eqnarray}
using the deformation parameters as the generator coordinates $q\equiv \beta,\gamma$. The GCM considers the nuclear wavefunction as a linear combination of (projected) reference states spanning some collective degree of freedom, $q$. Nevertheless, the initial reference states are not necessarily orthogonal. Thus, it is necessary to find a set of linearly independent wavefunctions, denoted as the natural basis. This is achieved by diagonalizing the overlap matrix defined as \cite{tesis}
\begin{eqnarray}
    \mathcal{N}^{\Gamma}_{qKq'K'}=\langle \phi(q) \vert {P}^{N} {P}^{Z} {P}^{J}_{KK'} \vert \phi(q') \rangle\,,
\end{eqnarray}
where $\Gamma\equiv(NZJM)$, and taking only the eigenstates $u^{\Gamma}_{\lambda;qK}$ with eigenvalues above a certain tolerance, $n^{\Gamma}_{\lambda}\geq\epsilon$. With this, the natural basis states are
\begin{eqnarray}
    \vert \Lambda^{\Gamma}_{\lambda} \rangle = \sum_{q'K'} \frac{u^{\Gamma}_{\lambda;q'K'}}{(n^{\Gamma}_{\lambda})^{1/2} } {P}^{N} {P}^{Z} {P}^{J}_{MK} \vert \phi(q) \rangle\,, \label{Natural}
\end{eqnarray}
and the GCM wavefunction in the natural space is
\begin{eqnarray}
    \vert \Psi^{\Gamma}_{\sigma,{\text{GCM}}} \rangle =\sum_{\lambda} G^{\Gamma}_{\sigma;\lambda} \vert \Lambda^{\Gamma}_{\lambda} \rangle\,,
\end{eqnarray}
with coefficients $G^{\Gamma}_{\sigma;\lambda}$ determined by solving the Hill-Wheeler-Griffin eigenvalue equation
\begin{eqnarray}
    \sum_{\lambda'} \langle \Lambda^{\Gamma}_{\lambda} \vert \mathcal{{H}}_{\text{eff}} \vert \Lambda^{\Gamma}_{\lambda'} \rangle G^{\Gamma}_{\sigma;\lambda'} = E^{\Gamma}_{\sigma} G^{\Gamma}_{\sigma;\lambda}\,. \label{HWG}
\end{eqnarray}

\begin{figure*}[t]
\includegraphics[width=0.8\linewidth]{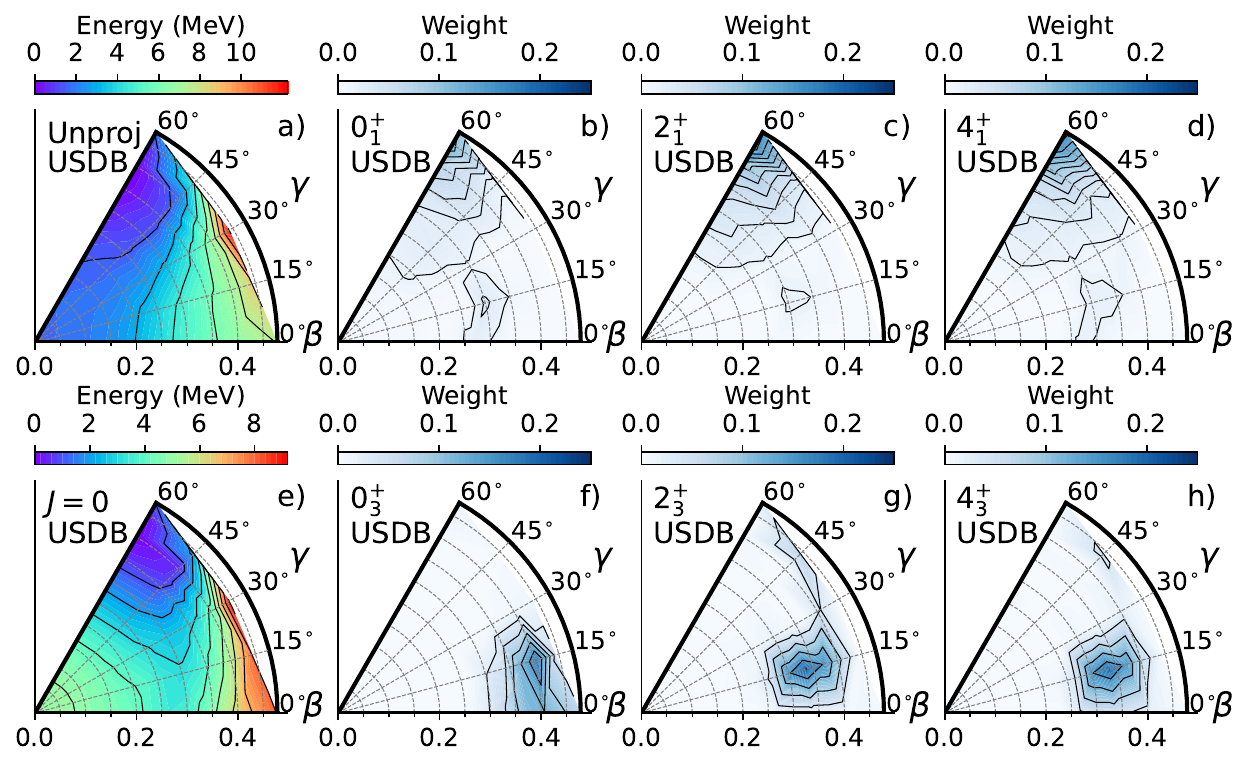}
\caption{{$^{28}$Si reference-state total energy surfaces and PGCM collective wavefunctions for the USDB interaction. The energy surfaces are unprojected (a) and projected to $N_v=6$, $Z_v=6$ and $J=0$ (e). The collective wavefunctions correspond to the lowest-energy $0^+$, $2^+$ and $4^+$ states with oblate (b-d) and prolate shape (f-h).}}
\label{fig:GCM_USDB}
\end{figure*}

These combined techniques are very useful for a direct exploration of quadrupole properties through the collective wavefunctions
\begin{eqnarray}
    \vert F^{\Gamma}_{\sigma} (q) \vert ^{2} = \vert \sum_{K\lambda} G^{\Gamma}_{\sigma;\lambda} u^{\Gamma}_{\lambda;qK}  \vert^{2}\,,
\end{eqnarray} 
which can be interpreted 
as the weight---not the probability, because the basis is not orthogonal---of each projected VAP wavefunction in the configuration-mixed state. Nonetheless, to extact firm conclusions the information given by the collective wavefunctions must be consistent with the intrinsic quadrupole moments obtained from $B(E2)$ transitions and quadrupole moments $Q_{\text{spec}}$ computed within the PGCM using Eq \ref{Qspec_def}.

\section{Results}
\label{sec:results}

We now discuss our findings for each of the main collective structures in $^{28}$Si: oblate, prolate and superdeformed states.  Whenever possible, we compare the results obtained with the variational approach and the  diagonalization, discussing the key strengths and weaknesses of each method. In addition, the corresponding outband $B(E2)$ transitions, including those between oblate and prolate states, are collected in Appendix~\ref{sec:outband}.

\subsection{Oblate band and $\beta$-vibration}
\label{sec:oblate}

\begin{table}[t]
\caption{{Quadrupole moments, $Q_{\text{spec}}$, for the lowest-energy oblate and prolate $2^+$ states in $^{28}$Si. Experimental values are compared to the PGCM and diagonalization (Diag.) results for the USDB, USDB-MOD and SDPF-NR* interactions.} 
}
\begin{tabular}{cccccccc}
\hline \hline
 & \multicolumn{7}{c}{$Q_{\text{spec}}$ ($e$~fm$^2$)} \\
 & \multicolumn{2}{c}{USDB} & \multicolumn{2}{c}{USDB-MOD} & \multicolumn{2}{c}{SDPF-NR*} & Experiment\\ \hline
 & \multicolumn{1}{c}{PGCM} & Diag. & \multicolumn{1}{c}{PGCM} & Diag. &  \multicolumn{1}{c}{PGCM} & Diag. & \\ \hline 
$2^+_{\text{obl}}$ & \multicolumn{1}{c}{18.5} & 18.5 & \multicolumn{1}{c}{19.4} & 19.2 &  \multicolumn{1}{c}{21.4} &  19.0 & $16\pm 3$~\cite{raghavan1989table}\\
$2^+_{\text{pro}}$ & \multicolumn{1}{c}{-5.0} & -7.9 & \multicolumn{1}{c}{-19.2} & -19.2  & \multicolumn{1}{c}{-16.9} &  -19.6  & \\ \hline \hline
\end{tabular}
\label{tab:moments}
\end{table}

\begin{table}[b]
\caption{$^{28}$Si ground-state energies relative to the $^{16}$O core obtained by diagonalization and the PGCM. 
}
\begin{tabular}{cccc}
\hline \hline
& \multicolumn{3}{c}{$0^+_\text{gs}$ energy (MeV)} \\
 & USDB        & USDB-MOD       & \textcolor{black}{SDPF-NR*} \\ \hline 
Diagonalization &  -135.9   &     -137.6           &  \textcolor{black}{-145.1}                    \\ \hline  
PGCM &    -135.4       &  -137.1 &   \textcolor{black}{-136.2}                     \\ \hline 
\end{tabular}
\label{tab:Energies_Si28}
\end{table}

We begin by studying the oblate band with the $^{28}$Si ground state as bandhead. Figure \ref{fig:Si28Spectrum_USDB} compares the experimental data (panel a) with the band structure obtained with the USDB interaction using the diagonalization and the PGCM results (panel~b). Both calculations agree very well with experiment, showing a clear rotational band with energies approximately proportional to $J(J+1)$ and strong $B(E2)$ transition strengths. In fact, the non-ideal behavior of the moment of inertia for the lowest-energy levels of the band is well captured by the USDB interaction.
\textcolor{black}{ Additionally, the $B(E2)$ transition strengths are comparable to the experimental ones, albeit slightly larger. The same is true for the USDB quadrupole moment \textcolor{black}{$Q_{\text{spec}}(2_1^+)=18.5\ e$~fm$^2$} listed in Table~\ref{tab:moments}.  
This is consistent with the $B(E2,2_1^+\xrightarrow{}0_{gs}^+)$ value, which is compatible with the experimental quadrupole moment $Q_{\text{spec}}(2_1^+)=16\pm3\ e$~fm$^2$ }\cite{raghavan1989table}. The electric quadrupole moments and transitions obtained with the PGCM agree very well with the exact results found by diagonalization. Moreover, the absolute energy of the ground state is only $\sim0.5$~MeV higher  than the one  obtained in the diagonalization, see Table \ref{tab:Energies_Si28}.

\begin{table}[b]
\caption{Occupation numbers for $^{28}$Si, $n(\text{orbital})$, of the $sd$-shell and summed $pf$-shell orbitals from diagonalization. The results are for the bandheads of the oblate and prolate bands obtained with the USDB, USDB-MOD and SDPF-NR* interactions.}
\begin{tabular}{lcccc}
\hline \hline
  & $n(0d_{5/2})$ & $n(1s_{1/2})$  & $n(0d_{3/2})$  & $n(pf)$   \\ \hline
USDB     &  &   &   &    \\ \hline
Oblate   & 9.32   & 1.43 & 1.25 &      \\ 
Oblate vibration & 9.41 & 1.46 & 1.13 \\
``Prolate''  & 7.84   & 1.64 & 2.52 &      \\  \hline 
USDB-MOD &  & &  &    \\ \hline 
Oblate   & 8.74   & 1.58 & 1.68 &      \\ 
Oblate vibration & 8.86 & 1.40 & 1.74 \\ 
Prolate  & 7.74   & 1.28 & 2.98 &      \\ \hline 
\textcolor{black}{SDPF-NR*}  &  &   &   &    \\ \hline
\textcolor{black}{Oblate}   & 7.87   & 1.52 & 1.68 & 0.93 \\ 
\textcolor{black}{Oblate vibration} & 8.43 & 1.29 & 1.47 & 0.81 \\
\textcolor{black}{Prolate}  & 6.79   & 1.50 & 2.71 & 1.00\\  \hline \hline
\end{tabular}
\label{tab:Occupations}
\end{table}

\begin{figure*}[t]
\centerline{%
\includegraphics[width=0.67\linewidth]{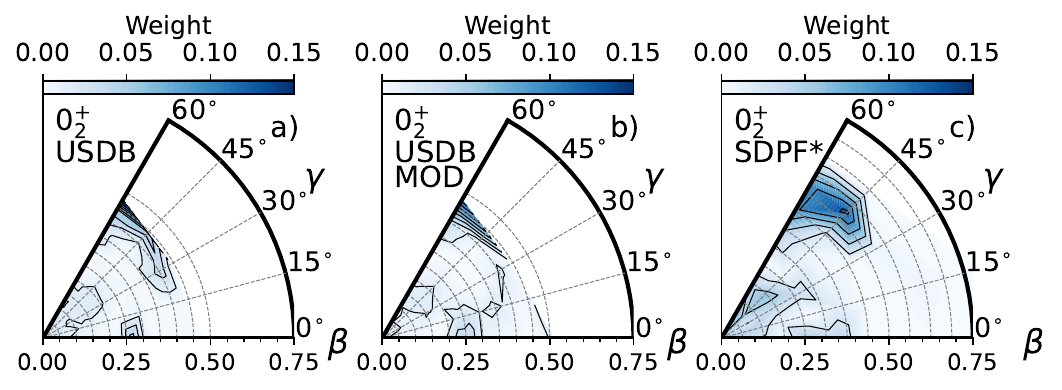}}
\caption{{$^{28}$Si PGCM collective wavefunctions of $0^+$ $\beta$-vibration states for the USDB (a), USDB-MOD (b)  and SDPF-NR* (c) interactions.}}
\label{fig:GCM_vibration}
\end{figure*}

\begin{figure*}[t]
\includegraphics[width=0.85\linewidth]{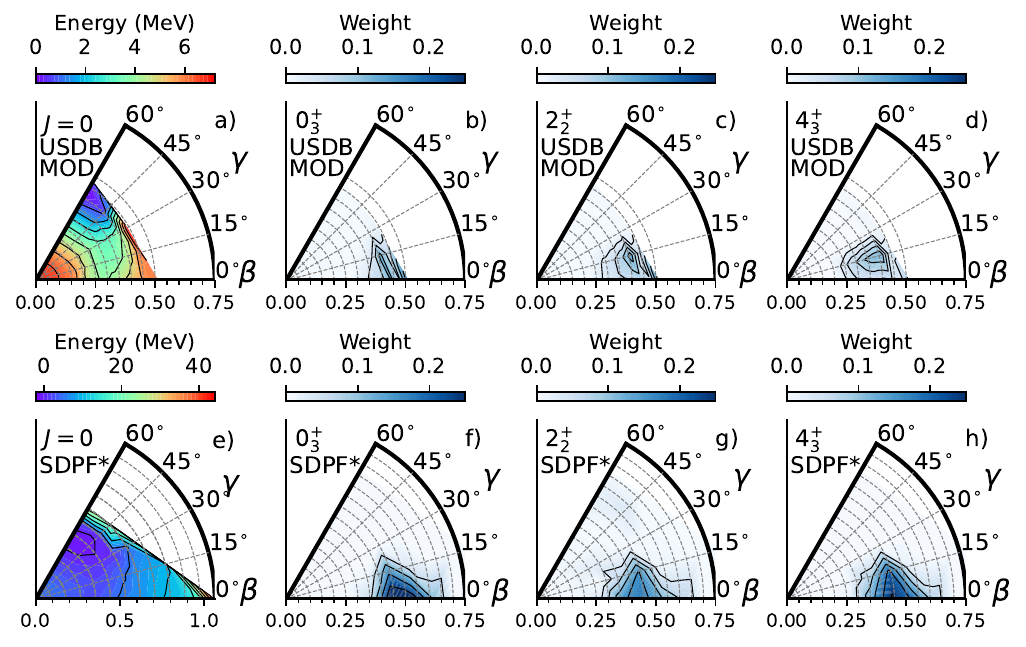}
\caption{{$^{28}$Si reference-state total energy surfaces and PGCM prolate collective wavefunctions. The energy surfaces,
projected to $N_v=Z_v=6$ and $J=0$, are calculated for the USDB-MOD (a) and SDPF-NR* (e) interactions. The collective
wavefunctions correspond to the lowest-energy $0^+$, $2^+$ and $4^+$ states with prolate shape obtained for USDB-MOD (b-d) and
SDPF-NR* (f-h).}}
\label{fig:GCM_prolate}
\end{figure*}

The PGCM provides additional insights on the structure of the oblate band. Figure~\ref{fig:GCM_USDB} shows the total energy surfaces of the reference states in a grid consisting of 63 VAP wavefunctions spanning the values of the deformation parameters \textcolor{black}{$0.000\leq\beta\lesssim0.478$} and $0^{\circ}\leq\gamma\leq60^{\circ}$ with spacing \textcolor{black}{$\delta_{\beta}\simeq0.035$} and $\delta_{\gamma}=15^{\circ}$.
The absolute minimum of both the unprojected surface (panel a) and the one projected to $N_v=6$, $Z_v=6$ and $J=0$ (panel e) corresponds to an oblate shape with \textcolor{black}{$\beta\simeq0.44$ ($\beta_2\simeq0.39$)}, which is consistent with the calculated quadrupole moments and $B(E2)$ strengths. 
Figures~\ref{fig:GCM_USDB} b), c) and d) also show the collective wavefunctions of the three lowest-energy states of the oblate rotational band including configuration mixing through the GCM as discussed in Sec.~\ref{sec:variational}. This is equivalent to the contributions of each projected VAP wavefunction to the 
$J=0^+_\text{gs}$, $2^+_1$, and $4^+_1$ mixed states. The common dominance of an oblate deformation across the three states confirms their identification in Fig.~\ref{fig:Si28Spectrum_USDB} as members of a well-established rotational band.

This picture is further supported by the occupation numbers obtained by the diagonalization. Table \ref{tab:Occupations} lists the values for the bandheads, but the occupation numbers are consistent across the band: $n(0d_{5/2})\approx9.3$, $n(1s_{1/2})\approx1.4$ and $n(0d_{3/2})\approx1.3$.
On the one hand, the $0d_{5/2}$ orbital presents the largest relative occupancy---$n(0d_{5/2})/[2(2j+1)]=0.78$---due to the $21\%$ contribution of closed $0d_{5/2}$ subshell.
On the other, the occupation numbers indicate that the ground state
is not entirely $0$p-$0$h as predicted by the quasi-SU(3) model. Nonetheless, the relative occupancy of the $1s_{1/2}$ orbital ($1.4/4=0.35$) is more than twice that of the $0d_{3/2}$ state ($1.3/8=0.16$). This difference is driven by quadrupole correlations, since in $^{28}$Si the two orbitals have very similar effective single-particle energies~\cite{Smirnova:2019yiq}.
The $n$p-$n$h excitations to the $0d_{3/2}$ orbital may account for the larger deformation obtained with the USDB interaction compared to the quasi-SU(3) value for the quadrupole moment listed in Table~\ref{tab:npnhconfig_analytical}.

\begin{figure*}[t]
\centerline{%
\includegraphics[width=0.67\linewidth]{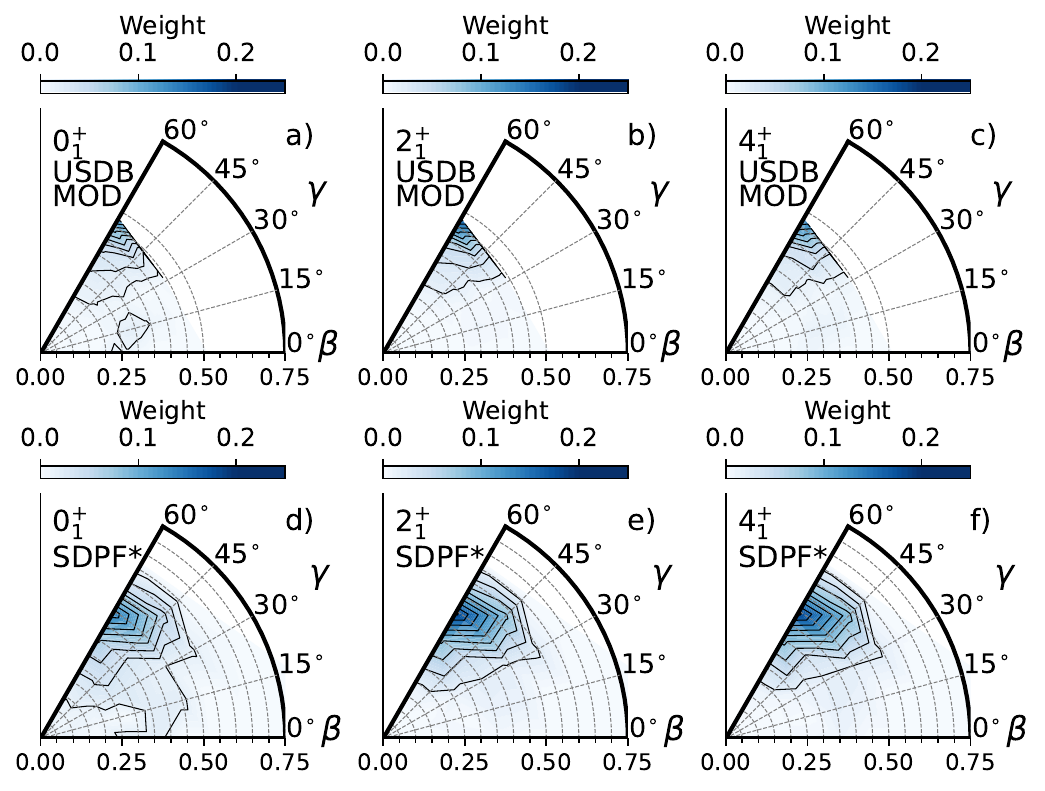}}
\caption{{Collective wavefunctions as in Fig.~\ref{fig:GCM_prolate}, but for the USDB-MOD (a-c) and SDPF-NR* (d-f) interactions and the lowest-energy oblate states.}}
\label{fig:GCM_oblate_others}
\end{figure*}

Finally, for the $\beta$-vibration band the theoretical  transition strengths in Fig.~\ref{fig:Si28Spectrum_USDB} also agree well with experiment. Our PGCM calculations support the interpretation of the first excited $0^+_2$ state, at an 
excitation energy of about $5$~MeV, as the bandhead of a $\beta$-vibration of the oblate shape. 
Figure~\ref{fig:GCM_vibration} a) shows a collective wavefunction that is almost identical to the ones for the ground state and for the associated rotational band, depicted in Figs.~\ref{fig:GCM_USDB} b), c) and d). Moreover, the results in Table \ref{tab:Occupations} indicate that the occupation numbers of the $0^+_2$ state and the ground state are very similar, further supporting the idea that this band has a $\beta$-vibration nature.

\subsection{Prolate band}
\label{sec:prolate}

In contrast with the oblate low-energy structure, a comparison between panels a) and b) of Fig.~\ref{fig:Si28Spectrum_USDB} highlights that the prolate band with bandhead at $\sim6.5$~MeV is not well described by the USDB interaction. Even though in the diagonalization the 
$0^+_3$, $2^+_3$ and $4^+_3$ states present an apparently rotational band spacing---albeit with bandhead $\sim1$ MeV higher than in experiment---, 
they are connected by much weaker transitions, \textcolor{black}{$B(E2;2^+_3\xrightarrow{}0^+_3)=34\ e^2$ fm$^4$ and $B(E2;4^+_3\xrightarrow{}2^+_3)=49\ e^2$ fm$^4$, than the measured $B(E2;4^+_3\xrightarrow{}2^+_3)=150\pm20\ e^2$ fm$^4$}. 
Consistently, the quadrupole moment of the $2^+_3$ state, listed in Table~\ref{tab:moments}, does not correspond to a well-developed prolate state either. 
The variational results in Fig.~\ref{fig:Si28Spectrum_USDB} and Table~\ref{tab:moments} also predict weak electric quadrupole transitions between these states. 
Further, the PGCM illustrates even more clearly the lack of a common structure: 
the collective wavefunctions in Figs.~\ref{fig:GCM_USDB}~f), g) and h) 
indicate that the $0^+_3$ bandhead presents a larger deformation than the $2^+_3$ and $4^+_3$ states.
\textcolor{black}{This loss of deformation is consistent with the quadrupole sum rule of the $0^+_3$ state, which yields an intrinsic quadrupole moment $Q_{0,\text{SR}}=71.7 e\ \text{fm}^2$ (similar to the experimental $Q_{0,t}=72\pm7  e\ \text{fm}^2$), but which is fragmented across several $2^+$ states.}
Therefore, even if, for convenience, we label these USDB states in Fig.~\ref{fig:Si28Spectrum_USDB} b) and Table~\ref{tab:Occupations} as ``prolate'', they do not show any feature of such deformed shape.  

Nonetheless, the results of the quasi-SU(3) analysis summarized in Table~\ref{tab:npnhconfig_analytical} suggest that a prolate structure can be accommodated within the $sd$ shell, provided that $2$p-$2$h and $4$p-$4$h excitations to the $0d_{3/2}$ orbital do not have to overcome a too large gap due to the orbital's single-particle energies.
In order to favor these excitations---while leaving the oblate rotational and vibrational bands unchanged---we propose a slightly modified interaction, denoted as USDB-MOD. We lower the single-particle energy for the $0d_{3/2}$ orbital by $1.2$~MeV, thus reducing the single-particle gap. The rest of the USDB-MOD interaction is the same as USDB. 

Figure \ref{fig:Si28Spectrum_USDB} c) and Table~\ref{tab:moments} show the results obtained with the new USDB-MOD interaction. Energy-wise the results are very similar to the ones for USDB, only with the prolate band appearing at $\sim 1$~MeV lower excitation energy. The main change concerns the electric quadrupole observables. 
\textcolor{black}{First, the $B(E2)$ transition strengths are now almost double compared to the USDB values. }
Second, the quadrupole moment of the $2^+_2$ state is consistently large, and supports a prolate shape. \textcolor{black}{Third, the $2^+_2$ amounts for an $80\%$-$90\%$ of the quadrupole sum rule value of the $0^+_3$ state}. Therefore, the results obtained with USDB-MOD suggest an actual prolate structure, unlike those predicted by USDB.
In turn, the oblate and vibrational bands remain almost unaltered.
The agreement between the diagonalization and the variational results is excellent. \textcolor{black}{However, the deformations we observe still differ moderately compared to experimental values, with our calculations indicating a larger deformation for the oblate band and a smaller deformation for the prolate band. Furthermore, the additional collectivity seems to be lost for the $6^+\xrightarrow{}4^+$ transition. This underscores the need for further improvement of the theory. }

The PGCM calculations also confirm the prolate nature of the band obtained with the USDB-MOD interaction.
First, at the mean-field level the total energy surface of USDB-MOD, Fig.~\ref{fig:GCM_prolate} a), shows a prolate minimum, in contrast with the corresponding surface of USDB, Fig.~\ref{fig:GCM_USDB}~e).
More importantly, Figs.~\ref{fig:GCM_prolate} b), c) and d) show that the $0^+_3$, $2^+_3$ and $4^+_3$ states share a common deformation across the band, although there is a minor loss in collectivity as $J$ increases. Further, Fig.~\ref{fig:GCM_oblate_others} shows that the ground-state band preserves its well-deformed character, and likewise Fig.~\ref{fig:GCM_vibration}~b) illustrates that the $\beta$~vibration is preserved as well.

Table~\ref{tab:Occupations} lists the occupation numbers of the prolate $0^+_3$ bandhead. The occupancy of the $0d_{3/2}$ orbitals for USDB-MOD is about $0.5$ nucleons higher than for USDB---equivalently, the combined $0d_{5/2}$ and $1s_{1/2}$ orbitals are less occupied by around $0.5$ nucleons. This change is consistent with the prediction of the quasi-SU(3) scheme. 
We note that very recently Ref.~\cite{lei2024multiplet}, using the variation pair condensate method, has also pointed out the possibility of taking into account the coexistence of oblate and prolate shapes in $^{28}$Si in an $sd$-shell calculation.

However, given the excellent performance of the USDB interaction across the $sd$ shell, we expect that the change introduced into USDB-MOD will translate into a lower-quality description of  nuclei neighboring $^{28}$Si. \textcolor{black}{Additionally, experimental data points out to a non-negligible occupation of the $pf$-shell orbitals for states at low energies~\cite{Nann:1982wxx}, highlighted by the presence of negative parity states at only 6 MeV of excitation energy. Finally}, the quasi-SU(3) scheme with an additional orbital discussed in Eq.~\eqref{eq:Singleorbit} and Fig.~\ref{fig:SU(3)orbitals} f) suggests that, if the $0d_{5/2}$-$1s_{1/2}$ doublet is complemented with $2$p-$2$h excitations to the $0f_{7/2}$ orbital---at a cost of overcoming the $sdpf$ shell gap---the associated deformation would be similar to the one achieved by the combination of the doublet with $4$p-$4$h contributions to the $0d_{3/2}$ orbital listed in Table~\ref{tab:npnhconfig_analytical}. 

Therefore, we explore further the structure 
of the prolate band expanding the configuration space to include the $pf$ shell. For this space we \textcolor{black}{start from} the SDPF-NR interaction \cite{SDPF}, \textcolor{black}{designed for neutron-rich silicon isotopes. We modify the interaction to accommodate cross-shell excitations, not permitted in the calculations of Ref.~\cite{SDPF}. To reproduce the $sdpf$ shell gap in $^{28}$Si, we take as reference the excitation energies of the negative-parity states of $^{28}$Si: 6.9 MeV ($3^-$), 8.4 MeV ($4^-$) and 8.9 MeV ($1^-$). By adjusting the $T=0$, $0d_{5/2}$-$0f_{7/2}$ and $0d_{5/2}$-$1p_{3/2}$,  monopole part of the interaction
we find that the negative-parity states at 6.9 MeV ($3^-$), 8.0 MeV ($1^-$) and 8.4 MeV ($4^-$), in good agreement with experiment. We name the modified interaction SDPF-NR*.}
Since the configuration space now involves two major harmonic oscillator shells, we take care of possible spurious center-of-mass contamination by adding to SDPF-NR the center-of-mass Hamiltonian, $\mathcal{H}_\text{cm}$, scaled by a factor $\lambda_\text{cm}=0.5$~\cite{Ca40}: $\mathcal{H}_\text{eff}'=\mathcal{H}_\text{eff}+\lambda_\text{cm}\mathcal{H}_\text{cm}$. Additionally, in the PGCM calculations we constrain $Q_{10}=0$ and $Q_{11}=0$.

\begin{figure}[t]
\centering
\includegraphics[width=1.0\linewidth]{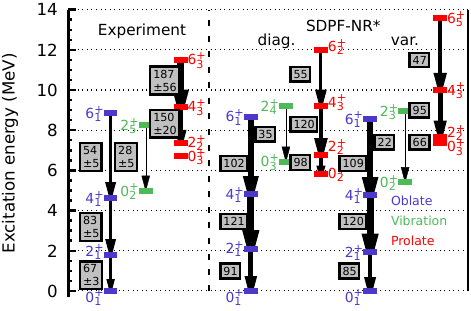}
\caption{{Same as Fig.~\ref{fig:Si28Spectrum_USDB}, but showing the experimental data (left) and the results obtained for the SDPF-NR* interaction with diagonalization (center) and the  PGCM (right).}}
\label{fig:Si28Spectrum_SDPF}
\end{figure}

Since we cannot perform the diagonalization in the full $sdpf$ space, we restrict the number of excitations from the $sd$ to the $pf$ shell to up to $4$p-$4$h. As we discuss in Sec.~\ref{sec:superdeformed}, this should be sufficient to capture the leading prolate configurations involving the $pf$ shell. 
\textcolor{black}{Figure~\ref{fig:Si28Spectrum_SDPF} (right side) shows the spectrum and electric quadrupole transitions for this truncated diagonalization. 
The $B(E2;4^+_3\xrightarrow{}2^+_2)$ transition is consistent (within 1.5$\sigma$) with the experimental value, supporting the interpretation as a prolate band.
In addition, the quadrupole moment for the $2^+_2$ state in Table~\ref{tab:moments} is also consistent with a prolate intrinsic shape and amounts to 90\% of the sum rule value. In contrast, the $B(E2;6^+_5\xrightarrow{}4^+_3)$ transition is still underestimated in our calculation.
Perhaps surprisingly, Table~\ref{tab:Occupations} indicates that the $0^+_2$ prolate bandhead only contains 1 nucleon in the $pf$ shell---excited from the $0d_{5/2}$-$1s_{1/2}$ orbitals---but this adds enough deformation to form a prolate rotational band.
Nevertheless, this indicates that $38 \%$ of the prolate state has a $2$p-$2$h $sdpf$ character, an important contribution to the wavefunction. The oblate and vibrational bands feature a similar occupation of the $pf$ shell, which leads to some  additional deformation compared to the USDB results.}


In addition to the usual insights, the PGCM allows us to consider the unrestricted $sdpf$ space. Figure~\ref{fig:GCM_prolate} e) hints to a prolate local minimum at the projected energy surface, and Figs.~\ref{fig:GCM_prolate} f), g) and h) show that the associated collective wavefunctions share a common prolate deformation across the band members. These results  confirm that the prolate band is well reproduced by the $sdpf$ calculation. At the same time, Fig.~\ref{fig:GCM_oblate_others} and Fig.~\ref{fig:GCM_vibration}~c) show that the ground-state and $\beta$-vibration bands keep their common,  well-defined structure in the $sdpf$ configuration space.

Regarding the reliability of the PGCM in this large configuration space, Fig.~\ref{fig:Si28Spectrum_SDPF}~(right side) shows that the oblate ground-state band is in almost perfect agreement with the truncated diagonalization, while the vibrational band is similar but appears $\sim 1$~MeV lower in excitation energy. In both cases, the $B(E2)$ values agree well with experiment.
In contrast, the prolate-band states appear at $\sim1.5$~MeV higher energy, and with slightly lower deformation than in the truncated diagonalization. The quadrupole moments of the oblate and prolate $2^+$ states in Table~\ref{tab:moments} also deviate somewhat from the diagonalization results. These differences could be related to the fact that the absolute energies in Table~\ref{tab:Energies_Si28} are \textcolor{black}{$\sim9$~MeV} higher for the PGCM than for the diagonalization. This difference points to non-captured correlations, such as pairing, which could be introduced as another generator coordinate for the configuration mixing; or the lack of dynamical correlations \cite{frosini2022multi}.  Nonetheless, since the disagreement between the diagonalization and the variational approach is mild, we expect the conclusions from the PGCM analysis to hold. 

\begin{figure}[t]
\centering
\includegraphics[width=1.0\linewidth]{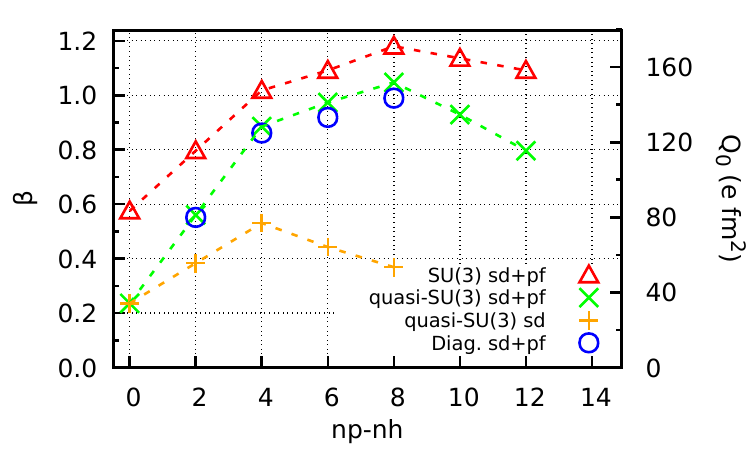}
\caption{{Deformation parameter $\beta$ (left axis) and the associated quadrupole moment $Q_0$ (right axis) for $^{28}$Si for fixed $n$p-$n$h structures in different  schemes: SU(3) $sd$+$pf$ shells (red triangles), quasi-SU(3) $0d_{5/2}$-$1s_{1/2}$ and $0f_{7/2}$-$1p_{3/2}$ pairs (green times symbols), and quasi-SU(3) $0d_{5/2}$-$1s_{1/2}$ plus the $0d_{3/2}$ orbital (orange crosses), compared to the fixed $n$p-$n$h shell-model results for SDPF-NR* (blue circles). For the latter, the space for $6$p-$6$h and $8$p-$8$h configurations is truncated to $sd$ + $0f_{7/2}$-$1p_{3/2}$.}}
\label{fig:sdpf_deformations}
\end{figure}

\subsection{Superdeformation}
\label{sec:superdeformed}

Reference~\cite{AMD} has proposed the appearance of superdeformed states in $^{28}$Si forming a rotational band with $0^+$ bandhead at $\sim13$~MeV excitation energy with \textcolor{black}{$\beta\simeq1$ ($\beta_2\simeq0.78$)}. In particular, this antisymmetrized molecular dynamics study assigns a $4$p-$4$h  $sdpf$ structure to the superdeformed band. This prediction motivated the search in Ref.~\cite{Si28_SD}, which however did not find evidence for superdeformation.

Here, we analyze the possibility for the existence of superdeformed states in  $^{28}$Si with the shell model.
First, we explore the deformations that can be achieved within the SU(3) schemes presented in Sec.~\ref{sec:SU3}. Figure~\ref{fig:sdpf_deformations} summarizes our results. The maximum deformation that can be achieved within the quasi-SU(3) $sd$-shell scheme presented in Table~\ref{tab:npnhconfig_analytical} is limited to \textcolor{black}{$\beta\lesssim0.53$} (orange crosses).
Thus, in order to build superdeformed states, we need to consider excitations into the $pf$ shell. 
In the pure SU(3) scheme in terms of $sdpf$ excitations, superdeformed shapes  with \textcolor{black}{$\beta\simeq1.0$ ($\beta_2\simeq0.78$)} appear at the level of $4$p-$4$h states (red triangles). Notably, higher $n$p-$n$h configurations offer mild gains in quadrupole correlations in comparison to the energy cost of exciting nucleons to the $pf$ shell. These quadrupole moments result from adding the contributions of filling the diagrams in Figs.~\ref{fig:SU(3)orbitals} a) and d) for $^{28}$Si. 
Likewise, the summed contributions of the diagrams in Figs.~\ref{fig:SU(3)orbitals} b) and e) offer the more realistic quasi-SU(3) values, where nucleons are excited across the lowest-energy pair of orbitals in each shell, from $0d_{5/2}$-$1s_{1/2}$ to $0f_{7/2}-1p_{3/2}$ (green times symbols). In spite of the reduced deformation compared to SU(3), the $4$p-$4$h configuration, with \textcolor{black}{$\beta\simeq0.9$ $(\beta_2\simeq0.72)$}, is superdeformed. Both SU(3) and quasi-SU(3) schemes predict the superdeformed band to be prolate.

To verify the validity of the analytical predictions, we perform fixed $n$p-$n$h diagonalizations in the $sdpf$ space using \textcolor{black}{SDPF-NR*}. 
The results, shown in Fig.~\ref{fig:sdpf_deformations} in blue circles, resemble closely the deformations of the quasi-SU(3) $sdpf$ scheme. Therefore, $4$p-$4$h excitations into the $pf$ shell could lead to superdeformed structures in $^{28}$Si.

\begin{figure}[t]
\centerline{%
\includegraphics[width=1.0\linewidth]{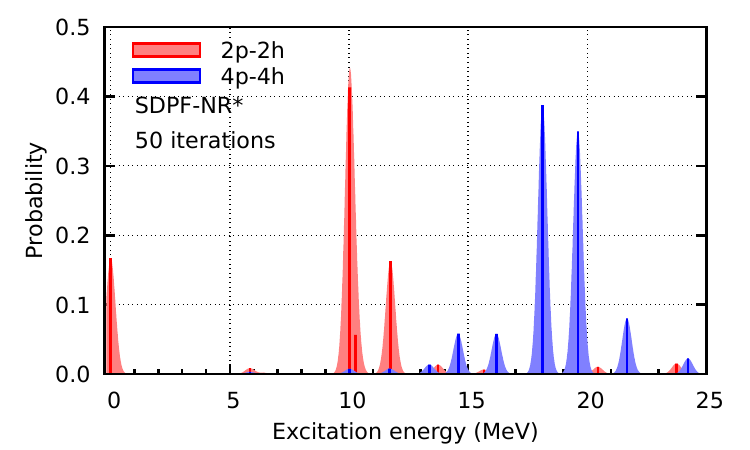}}
\caption{{Probability of finding the lowest-energy state in $^{28}$Si with fixed $2$p-$2$h and $4$p-$4$h $sdpf$ configuration as part of $0^+$ states of the $sdpf$ space, for the SDPF-NR*
interaction. Each $sdpf$ state is convoluted with a Gaussian of width $200$ keV.}}
\label{fig:Lanczos_superdeformed}
\end{figure}

In order to explore the energies associated with these $n$p-$n$h configurations, one should diagonalize the \textcolor{black}{SDPF-NR*} interaction in the full $sdpf$ space. In our case, to manage the dimension of the configuration space we restrict the number of nucleons in the $pf$ orbitals to $n(pf)\leq 4$. According to Fig.~\ref{fig:sdpf_deformations} this truncation should capture the main highly-deformed structures of $^{28}$Si. In this truncated space we use the Lanczos strength function method to expand the $\vert 0^{+}_\text{$n$p-$n$h} \rangle$ states in terms of the eigenstates of the $sdpf$ space:
\begin{eqnarray}
    \vert 0^{+}_{n\text{p}-n\text{h}} \rangle= \sum_{i} a_i \vert 0^{+}_{i} \rangle\,,
\end{eqnarray}
where  $a_i$ are the amplitudes of the expansion. Figure~\ref{fig:Lanczos_superdeformed} shows the Lanczos strength functions for the lowest-energy collective $0^+$ states with fixed $2$p-$2$h and $4$p-$4$h $sdpf$ configuration. These states contribute mostly to $sdpf$ excited states with energies around \textcolor{black}{$10$~MeV and $18$-$20$~MeV}, respectively.
In fact, in contrast to the $2$p-$2$h case, Fig.~\ref{fig:Lanczos_superdeformed} highlights that the superdeformed $4$p-$4$h structure does not contribute significantly to any state below  \textcolor{black}{$15$~MeV} in excitation energy. 

\begin{figure}[t]
\centerline{%
\includegraphics[width=0.9\linewidth]{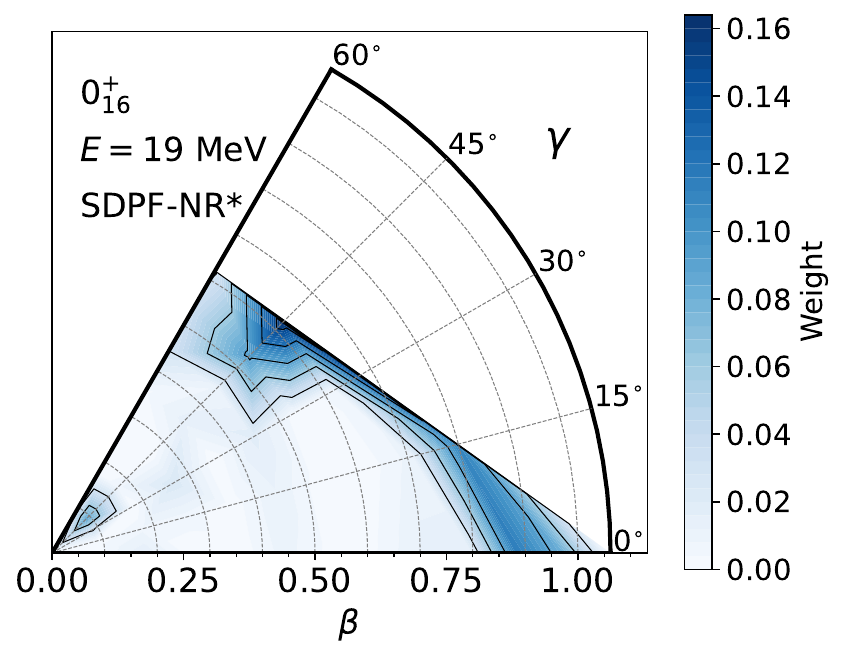}}
\caption{{PGCM collective wavefunction of the lowest-energy candidate superdeformed state $(0^+_{16})$ for $^{28}$Si, obtained with the SDPF-NR* interaction.}}
\label{Possible_superdeformed}
\end{figure}

The PGCM allows us to perform a complementary study of the possible superdeformation in $^{28}$Si considering the full $sdpf$ configuration space. 
For the basis, we chose a grid of 42 VAP wavefunctions with spacing \textcolor{black}{$\delta_{\beta}=0.089$} up to a maximum \textcolor{black}{$\beta=1.06$} and $\delta_{\gamma}=15^{\circ}$.
\textcolor{black}{Consistently with the diagonalization results, our PGCM calculations feature the lowest-energy state that could be associated with a superdeformed $0^+$ bandhead at $19$~MeV excitation energy. Figure~\ref{Possible_superdeformed} shows the collective wavefunction for this state, with $\beta\simeq0.6$ $(\beta_2\simeq0.51)$. This state has on average $3$ nucleons in the $pf$ shell. Even though this high-energy state obtained with the variational approach may not be accurate enough to correspond to the exact solution~\cite{GCM_optimization}, the consistency between the diagonalization and the PGCM results indicate the appearance of superdeformed states in $^{28}$Si at 18-20~MeV excitation energy}.

Finally, we note that while this conclusion may appear in tension with the prediction of Ref.~\cite{AMD}, our shell-model configuration space cannot accommodate cluster structures such as the $^{24}$Mg+$\alpha$ suggested for the $^{28}$Si superdeformed band~\cite{AMD}.

\section{Summary and outlook}
\label{sec:summary}

In this study, we have analyzed the shape coexistence of $^{28}$Si combining different approaches: the analytical quasi-SU(3) model; standard shell-model diagonalizations, and the variational PGCM.

We have found that the oblate ground-state rotational band and its associated $\beta$-vibration band are well described with the gold-standard USDB interaction in the $sd$ shell. However, USDB fails to reproduce the experimental prolate rotational band. These conclusions are supported by the comparison between experimental and calculated electromagnetic moments and transitions via diagonalization, and by the analysis of collective wavefunctions using the PGCM. Likewise, both the diagonalization and the PGCM results show that the prolate band can be described in the $sd$ shell, at the cost of lowering the single-particle energy of the $0d_{3/2}$ orbital by about 1~MeV. We name the resulting interaction USDB-MOD. 
Alternatively, the band can be reproduced by considering an extended configuration space including the $pf$ shell \textcolor{black}{with a modified SDPF-NR interaction that can accommodate cross-shell excitations}.
In both cases, the oblate structures of $^{28}$Si are still well reproduced. The quasi-SU(3) model explains qualitatively the key role of the nucleon excitations into the $0d_{3/2}$ and $pf$-shell orbitals in order to build the prolate deformation. 

Additionally, we have explored the possibility of superdeformation in $^{28}$Si, predicted in previous works but not found experimentally so far~\cite{Si28_SD}. According to the quasi-SU(3) scheme, such extreme shape requires the promotion of at least $4$p-$4$h nucleons to the $pf$ shell. 
\textcolor{black}{Both our diagonalizations in a truncated $sdpf$ configuration space, and our variational PGCM calculations considering the full space indicate that superdeformed structures appear at $\sim18$-$20$ MeV of excitation energy.}

In most cases, the variational PGCM  results are in excellent agreement with the diagonalization. However, in the largest $sdpf$ space, the difference in absolute energies between the diagonalization and the PGCM can reach $\sim5$\%. These results could be improved by considering additional generator coordinates, such as the isovector or isoscalar pairing \cite{hinohara2014proton,Bally:2019miu,sanchez2021variational}, or by including dynamical correlations---which are more relevant in larger spaces---for instance via multi-reference many-body perturbation theory~\cite{frosini2022multi}.

The shape coexistence in $^{28}$Si can also be studied within our shell-model framework using {\it ab initio} interactions~\cite{Jansen:2014qxa,Stroberg:2019mxo,Smirnova:2019yiq}. For instance, those derived with the VS-IMSRG~\cite{Miyagi:2020ltz} have been recently applied to study $sd$-shell~\cite{Heylen:2020cco,Jokiniemi:2021qqg,Zanon:2023qem,Li:2023bel} and $sdpf$-shell nuclei~\cite{Kitamura:2022lem,Gray:2023bfa,Lubna:2024bko,Yuan:2024asg}. It would be actually quite interesting to explore whether the VS-IMSRG Hamiltonian for $^{28}$Si, which is tailored just for this nucleus, describes both the oblate and prolate bands within the $sd$-shell---as suggested by the USDB-MOD interaction. Or if, in contrast, the $sd$ valence-space correlations that drive the low-energy nuclear structure of $^{28}$Si are not sufficient to capture the prolate band, as we have found with USDB.     

While our results agree very well with experiment in most cases, some differences remain. For example, the collectivity of the ground-state oblate band is \textcolor{black}{slightly} overpredicted in all our calculations, \textcolor{black}{while the prolate band is underpredicted, especially when the $6^+$ state is involved}. We have also predicted so far unknown inband and outband electric quadrupole $B(E2)$ values, that could be tested in forthcoming measurements. As an outlook, we also plan to expand our study of $^{28}$Si by investigating the different octupole bands known experimentaly~\cite{Si28exp} within our shell-model framework. In addition, we aim to explore the recently-measured hexadecupole deformation in this nucleus~\cite{Gupta:2023cvv}.

More generally, our results show the powerful predictive power of complementing standard shell-model diagonalizations with the variational PGCM to study shape coexistence in nuclei.  Similar analyses could be carried out in other $N=Z$ nuclei in the $sd$ shell, such as 
$^{24}$Mg or $^{32}$S, or in neutron-rich Si isotopes such as $^{30-42}$Si, which can also exhibit shape coexistence~\cite{dowieMg24,Caurier:1998zz,SDPF,rodriguez2000S32,Rotaru:2012dy,Caurier:2013aoa,Han:2017jkk,Gupta:2023cvv}. Eventually, the PGCM enables us to address nuclei across the nuclear chart.    

\acknowledgments

We are grateful to Nobuo Hinohara, Paul Garrett, Frederic Nowacki and Alfredo Poves for very useful discussions.
This work is financially supported by 
MCIN/AEI/10.13039/5011
00011033 from the following grants: PID2020-118758GB-I00, PID2021-127890NB-I00, RYC-2017-22781 and RYC2018-026072 through the “Ram\'on
y Cajal” program funded by FSE “El FSE invierte en tu futuro”, CNS2022-135529 and CNS2022-135716 funded by the
“European Union NextGenerationEU/PRTR”, and CEX2019-000918-M to the “Unit of Excellence Mar\'ia de Maeztu 2020-2023” award to the Institute of Cosmos Sciences; and by the Generalitat de Catalunya, grant 2021SGR01095.

\bibliography{references}

\appendix
\section{Outband transitions}
\label{sec:outband}
Here we collect  $B(E2)$ transition strengths, including the inband transitions given in Fig.~\ref{fig:Si28Spectrum_USDB} and Fig.~\ref{fig:Si28Spectrum_SDPF}, and outband transitions, this is, those connecting states that belong to different bands according to the scheme shown in Fig.~\ref{fig:Si28Spectrum_USDB}. Table~\ref{tab:Transitions_all} lists all these results obtained by diagonalization and the PGCM, and compares them to the experimental values when available.

For outband $B(E2)$s, the PGCM and diagonalization results agree well for both the USDB and USDB-MOD interactions. However, the two methods present larger discrepancies for the SDPF interaction,  especially for the transitions connecting the prolate and vibrational bands, where the PGCM predicts larger $B(E2)$ values.
In general, the calculated outband transitions are in reasonable agreement with the measured ones\textcolor{black}{, except for the $B(E2,2_\text{obl}\xrightarrow{}0_\text{pro})$ and $B(E2,2_\text{obl}\xrightarrow{}0_\text{vib})$ from the SDPF-NR* diagonalization, which overpredicts the data. On the other hand, it is only the latter calculation that gives a good agreement with the experimental  $B(E2,4_\text{obl}\xrightarrow{}2_\text{vib})$ value}. 

\begin{table*}[]
\caption{\textcolor{black}{$B(E2)$ transition strength values involving initial ($J^+_{i}$) and final ($J^+_{f}$) $0^+$, $2^+$ and $4^+$ states in the oblate, $\beta$-vibration and prolate bands in $^{28}$Si, labeled as in Fig.~\ref{fig:Si28Spectrum_USDB}. Theoretical PGCM and diagonalization results for the USDB, USDB-MOD and SDPF-NR* interactions are compared to experimental data} \cite{Si28exp}.
}
\begin{tabular}{ccccccccc}
\hline \hline
 &  &   & \multicolumn{2}{c}{USDB} & \multicolumn{2}{c}{USDB-MOD} & \multicolumn{2}{c}{SDPF-NR*} \\ \hline
 &  & Experiment & PGCM & Diag. & PGCM & Diag. & PGCM & Diag. \\ \hline
\multicolumn{1}{c}{$J^+_{i}$} & \multicolumn{1}{c}{$J^+_{f}$} & \multicolumn{7}{c}{$B(E2)$ ($e^2$ fm$^4$)}  \\ \hline
2$_\text{obl}$ & 0$_\text{obl}$ & 67$\pm$2 & 78.3 & 78.3 & 87.7 & 86.9 & 84.7& 91.3\\
4$_\text{obl}$ & 2$_\text{obl}$ & 70$\pm$7 & 110 & 110 & 117 & 117 & 120& 121\\
2$_\text{vib}$ & 0$_\text{vib}$ &   28$\pm$7 & 18.2 & 21.8 & 26.3 & 25.1 & 22.1& 35.3\\
2$_\text{pro}$ & 0$_\text{pro}$ &    & 12.9 & 34.1 & 73.3& 73.4 & 65.8& 98.4\\
4$_\text{pro}$ & 2$_\text{pro}$ & 147$\pm$25 & 58.0 & 49.3 & 97.1 & 96.3 & 95.1& 120\\ \hline
2$_\text{obl}$ & 0$_\text{vib}$ & 
8.7$\pm$1.6 & 13.1 & 12.85 & 12.1 & 11.5 & 0.04& 39.2\\
2$_\text{obl}$ & 0$_\text{pro}$ & 0.27$\pm$0.02   & 0.17 & 0.58 & 0.27 & 0.14 & 0.04 & 17.7 \\
2$_\text{vib}$ & 0$_\text{obl}$ &   0.15$\pm$0.05 & 0.26 & 0.04 & 0.08 & 0.11 & 0.00& 0.12\\ 
2$_\text{vib}$ & 0$_\text{pro}$ &    & 43.7 & 39.2 & 26.6 & 24.0 & 31.5& 1.27\\ 
2$_\text{pro}$ & 0$_\text{obl}$ &    & 0.15  &  0.74 &  0.47 &  0.20 & 1.06& 0.02\\ 
2$_\text{pro}$ & 0$_\text{vib}$ &    & 5.47 & 9.58  & 23.4  & 16.0   & 18.9& 1.02\\ 
4$_\text{obl}$ & 2$_\text{vib}$ & 7.2$\pm$2.7   & 2.66   & 1.08 & 3.20 & 2.60  & 1.73& 10.75\\ 
4$_\text{obl}$ & 2$_\text{pro}$ &    &  2.91 & 2.68  & 0.74   &  0.91 & 1.67& 6.95\\  
4$_\text{pro}$ & 2$_\text{obl}$ &    & 0.30   & 0.11 & 1.27  & 0.89  & 6.39& 0.91\\ 
4$_\text{pro}$ & 2$_\text{vib}$ &    & 6.15  & 7.71 & 1.50  & 0.81  & 6.87& 0.10\\  \hline
2$_\text{vib}$ & 2$_\text{obl}$ &    & 0.27 & 0.14  & 6.83 & 6.20  & 4.25& 5.75\\ 
2$_\text{vib}$ & 2$_\text{pro}$ &    & 23.7  & 10.0  & 11.0   & 7.62  & 23.1& 5.94\\
2$_\text{pro}$ & 2$_\text{obl}$ &    & 1.18  & 6.04  & 6.45  & 7.10  & 12.0& 15.4\\
4$_\text{pro}$ & 4$_\text{obl}$ &    &  2.00   &  2.18  & 0.52   & 0.83  & 1.72 & 2.64\\  \hline \hline
\end{tabular}
\label{tab:Transitions_all}
\end{table*}

\end{document}